\documentclass[10pt,preprint]{aastex}
\usepackage{color}
\begin{document}

\title{Testing the Viewing Angle Hypothesis for Short GRB with LIGO events} 
\author{David Eichler$^{*}$}
\altaffiltext{*}{Dept. of Physics, Ben-Gurion University, Be'er-Sheva 84105, Israel}

\begin{abstract}
It has been suggested that short gamma ray bursts (GRBs) have shorter or undetectable spectral lags than long GRB because in the former, the observer's line of sight makes a larger angle with the GRB jet axis than for the latter. It is proposed that  simultaneous gravitational wave - short GRB events could provide a simple test of this hypothesis. { Multimessenger astronomy eventually may test whether event horizons are a necessary ingredient for GRB.}
\end{abstract}

\section{Introduction}

It has been  suggested in previous work  ( Eichler, Guetta and Manis, 2009)   that short GRBs  are typically viewed from a larger ``offset'' angle $\theta$ from the GRB jet velocity vector than are long GRB .  
If short GRBs are from coalescing neutron stars, as opposed to long bursts, which are 
from collapsing stellar cores inside massive WR stars, then it is hardly surprising that, on the average, the short GRBs should  be viewed from a larger offset angle than long GRBs, because the coalescing neutron stars have no stellar envelope that might, at least in the early stages, obscure the jet from observers with large offset angles.    Moreover, the envelope of collapsing cores may collimate GRB jets, and GRBs  { that pass through them may emerge with} smaller opening angles (and perhaps even different terminal Lorentz factors)  than those from coalescing neutron stars. When there is no envelope, the opening angle may be wider.  It has been widely argued that the envelopes  of long GRB progenitors require much more than a second - the  characteristic length of short bursts -  for the jet to bore a hole through them, so a short GRB inside massive stellar envelope would in any case be unobservable.\footnote{It should be noted, however, that the back end of an ultrarelativistic jet can push on the working surface at the front for longer than the duration of the jet because the front end advances forward during the interaction.}

The question remains of course:  Why,   physically,  do  GRBs from coalescing neutron stars seem shorter than those from collapsing cores? Although the central engine of a coalescing neutron star might be somewhat different from that resulting from a collapsing core,  the coalescing neutron star binary should if anything have more angular momentum and, arguably, last longer.  In contrast, if the timescale of GRB is associated with the hydrodynamics in the {\it environment} surrounding the central engine,  (e.g. the  entry and/or exit of baryons into and/or out of  the path of the GRB jet, or the time scale over which they are accelerated),  then one might expect coalescing neutron stars, where the matter distribution in  the surrounding environment probably has a much smaller spatial scale than the envelope of a WR star, to have a correspondingly shorter time scale. 

The apparent duration of a GRB need not be the same as that of the central engine. Kinematic effects such as motion and oblique scattering can lengthen the observed duration in the same manner that it softens the spectrum, as discussed below.\footnote{To see this easily think of the $\gamma$-radiation as a wave and note that the number of wave crests is the same before and after scattering}  Acceleration of the scattering material  beyond the Lorentz factor $1/\theta$ can cause the GRB to disappear from view.

A further complication arises: Short GRBs frequently have long, soft X-ray tails (sometimes called extended emission)(Mazets et al, 2002; Norris and Bonnel, 2006).  These tails are softer even than typical X-ray flashes, which are soft versions of long GRBs, and may themselves be long GRB/X-ray flashes observed at large offset angles.  The X-ray tails last as long as very long GRBs. Thus, the difference between `so called ``short'' and ``long'' GRB  really  refers to the duration of the hard $\gamma$- ray emission, not necessarily to the activity of the central engine. 
However,  another objective difference between long and short GRB is that long ones have measurable  spectral lags (Norris  et al 1996) - spectral evolution from hard to soft over several seconds - while short GRB have,  if any, spectral lags that are too short to measure.

 In attempting an answer to the above puzzles, it is worth noting two fundamentally surprising aspects of GRBs: That the quantity of baryons needed to power the GRB is enough to obscure it. Moreover, the luminosity of a GRB is so high that the radiation pressure ought to drag a highly opaque baryonic wind  out with the photons, in which they would adiabatically cool well before they could be observed.  The second problem would seem to prevent any scenario, motivated by the first problem, in which the baryons could be tidily sequestered so as not to block the line of sight. While a baryon-free zone can be created on horizon-threading field lines  after the formation of the black hole,  one still expects that this zone has a length of at most cT where T is the time since the formation of the event horizon,   and  ahead of that zone there be  an optically thick  wind of baryons, which probably commenced before the event horizon formed.
 
 Another curious distinction between short and long GRB is that long GRB typically obey the Amati relation (Amati et al, 2002, 2008, 2009: Amati, 2006),  $W_{iso}/10^{52}$ ergs $\simeq (h\nu_{peak}/100 KeV)^2$, (where $\nu_{peak}$ is the frequency at which the $\gamma$-ray spectrum peaks)  whereas short GRB, which have somewhat harder spectra than long GRB,  are typically  less luminous than predicted by the Amati relation for long GRB. { They may, however, have a similar relation with a different normalization (Ghirlanda et al, 2009).} More curious is that the soft X-ray tails of the short bursts {\it do obey}  the Amati relation for long GRB and X-ray flashes. {\bf In section 2 the viewing angle hypothesis interpretation of the above curiosities is outlined. In section 3 this interpretation is compared with very recent observations.}

 \section{The Viewing Angle Explanation for Soft GRB} 
 It has been suggested that all of the above can be understood if the $\gamma$-rays from GRB are first emitted outward in the baryon-pure zone (Levinson and Eichler, 1993) and are scattered by the baryons ahead of them (Eichler, 2014; Eichler and Manis, 2007, 2008;  Eichler, Guetta and Manis, 2009). { These photons could of course be emitted in flight  by pairs and perhaps even via their annihilation, provided that the pair-photon fluid is moving at a higher Lorentz factor than the slower material at the front of the jet. The reverse shock behind the slower matter at the front could in principle emit non-thermal radiation. For convenience, however, we take the fast part of the ``jet'' to be radially { directed} photons.} If the scattering surface is being accelerated by the radiation pressure of the photons, then the observed energy of the photons $E''\equiv h\nu''$ is 
 \begin{equation}
 E'' = D(\theta)E'=D(\theta){\cal D}E =\left([1-\beta]/[1-\beta cos\theta]\right)E 
 \end{equation}
 where 
 $E\equiv h\nu$ is the emitted photon energy in the frame of the central engine, $E'$ is the photon's energy in the frame of the scattering surface,   $D(\theta)$ is the Doppler factor $1/\Gamma[1-\beta cos\theta]$, ${\cal D} = 1/\Gamma(1+\beta)$,  and where $\beta c$ is the velocity of the scattering surface and $\Gamma$ is its Lorentz factor. 
 Consider now what the observer sees as $\Gamma$ increases, assuming the emission from the central engine is steady. Assume $\Gamma \gg 1$ and $\beta \simeq 1$, so that $D=1/\Gamma\left[(1-\beta) +\beta \left(1 - cos(\theta)\right)\right] \simeq 1/\Gamma\left[(1-\beta) + \left(1-\cos(\theta)\right)\right]$ and hence $E''\simeq \left(1+ \left[{1-cos\theta\over(1-\beta)}\right] \right)^{-1}E=E/[1+(\gamma\theta)^2]$.  As long as $\Gamma \le 1/\theta$ [i.e $\beta\le cos\theta$], the GRB luminosity rises with $\Gamma$ if unobscured by baryons, or is obscured by them, while the peak frequency does not change much. 
 As soon as $\Gamma$ accelerates beyond $1/\theta$, the emission  appears, even if the scattering surface is opaque,  because it is scattered  backward in the frame of the scattering surface, but, as it accelerates further,  the observed radiation declines  and softens.\footnote{ The statement is not always exactly true because of possible curvature of the scattering surface,  perhaps exacerbated by Raleigh-Taylor instabilities. It could be, if emission perpendicular to the jet axis doesn't escape, that escaping emission has to escape backwards in the frame of the scattering material and  that the Lorentz factor needed to beam it at the observer  needs to be  higher than $1/\theta$.} {\it Thus, the spectral evolution timescale (spectral lag time) of the GRB may be of the order of the acceleration time $\tau_{acc}(\Gamma)\equiv dln\Gamma/dt$ at the $\Gamma=1/\theta$ peak, or, if the radiation is obscured at this peak, whenever it first becomes visible.} It is straightforward to show, for a given radiation pressure and baryonic shell rest mass associated with the scattering surface, that the acceleration time $\tau_{acc}$ in the frame of the central engine scales as $\tau_{acc}(\Gamma)\propto \Gamma^{5}$ { (Eichler and Manis, 2007, 2008)}.  In observer time $\tau_{obs}$, which is compressed by $1-\beta  cos\theta \sim \Gamma^{-2}$  relative to  time in the frame of the central engine, the acceleration time  scales as $\tau_{obs}\propto \Gamma^{3}$. Moreover $\tau_{acc}$ correlates inversely with the GRB luminosity,  all other things being equal.   Such an inverse correlation is observed for the spectral lags of {\it long} GRBs (Norris et al, 2000; Gehrels, et. al, 2006), suggesting that their spectral lag times are also determined by an acceleration timescale. Further support for this hypothesis is that the spectral evolution  - pulse duration $\Delta t \propto (h\nu)^{-0.4}$ - is well explained quantitatively by attributing it to acceleration (Eichler and Manis, 2007, 2008).  However, the normalization for the correlation need not be the same for short as for long GRB.  In the case of the short GRB, the  baryonic shells that are accelerated  (which may originate from the neutron stars) are probably much closer to the source than for long GRB (where the baryons may originate from the stellar envelope), and  acceleration timescales  of the short GRBs may therefore perhaps be typically  too small to be resolved observationally. (The duration of the short GRB, which {\it is} resolvable, would then have to be attributed not to the acceleration timescale, but perhaps rather to the timescale of the baryon loading of the jet.  For example, a parade of baryonic blobs may continue to cross  or remain in the path of the jet for a longer time than the acceleration timescale of each blob.)
 
   Now assume that the short GRB can be viewed at an earlier stage and at a larger  offset angle. It has   shorter spectral lags, because $\tau_{acc}$ is smaller, and a softer X-ray tail over the latter stages of central engine activity, because $1-cos\theta$ is larger. 
   
   Because the observed photon energy declines with observer time, the fact that short GRB can be observed at earlier stages suggests that they would have somewhat harder spectra. At the moment $\Gamma=1/\theta$, the observed frequency $E''$ is $E/(1+\beta)\simeq E/2$. The typical value of $E_{peak}\equiv h\nu_{peak}$  can be estimated from the maximum peak energy ever observed for GRB, 2 to 4 MeV, where I assume these values are observed by the rare observer who looks right down the axis of the jet ($\theta \simeq 0$).  { This is consistent with the observation that short, hard GRB have spectral peaks between 1 and 2 MeV}.  The fact that long GRB typically have  peak energies much less than this - typically about 200 to 300 keV -  suggests that they may attain $\Gamma=1/\theta$ while still obscured by baryons in (or originating from) the envelope of the host star.  In this case, they would become visible to the observer only after $\Gamma> 1/\theta$ with correspondingly longer acceleration times. Short GRB, assuming they are less obscured, may be observable at an earlier stage,  closer to the point where $\Gamma=1/\theta$, and thus have harder spectra.
    
    The above scenario can be further constrained by  the Amati relation,   obeyed by X-ray tails of short GRB but not by the early hard stage. Within the framework of the above model, the Amati relation is derived from the premise that the opening angle of the jet, believed to be several degrees on the basis of observed jet breaks,  is larger than $1/\Gamma $, where $\Gamma$ is the Lorentz factor of the GRB jet, believed to be $\gtrsim 10^2$. If the material between the last scattering and the observer is optically thick, however, then the last scattering could not have been of material moving directly at the observer.  Rather, it must come from some material whose velocity vector makes an angle $\theta \ge \theta_{min}\ge 1/\Gamma$ with the observer's line of sight. That way, if it scatters into the backward hemisphere in the frame of the scattering surface, it may, without ever again making contact with the scattering surface,  be directed at the observer by relativistic kinematics. Here $\theta_{min}$  is the minimum angular difference between the line of sight and the observable part of the jet, which we assume has radial motion everywhere. If the jet were an infinitely narrow ``pencil'' beam, with a single value of $\theta$, the relation between $W_{iso}= dW/4\pi d\Omega$, the total energy per solid angle, and $E''$ would be
    \begin{equation}
    W_{iso}= D^3  dW'/4\pi d\Omega ={\cal D}^3 (E''/E )^3 dW'/4\pi d\Omega,
    \label{pencil}
    \end{equation} 
However, precisely because the luminosity drops off so fast with $E''$, relatively few GRB would be observed at large angles. But if the jet is shielded by an intervening optically thick cloak of baryons,  then emission at $\theta=0$ could also not be observed. Now if the minimum angular separation between the line of sight and the detected jet material $\theta_{min}$ is small, relative to the opening angle of the jet, but nevertheless finite, then the solid angle of jet that contributes to $W_{iso}$ is proportional to $(1-cos\theta_{min})$ which is approximately { $\Gamma E/E''$.} In this case, where the jet is extended, the distribution of collected energy    (isotropic equivalent energy) $W_{iso}$ in   GRB photons of emitted energy E for an isotropic collection of observers - each seeing his own $E''$ -  is
\begin{equation}
W_{iso} \propto (E''/E)^2
\end{equation}
which is the Amati relation.
The considerable scatter in the Amati  relation - about two orders of magnitude in  $W_{iso}$ -  may  be attributed to the scatter in the other parameters of GRB, but the range of observable values for $W_{iso}$ extends over 5 orders of magnitude,  from $10^{49}$ ergs to $10^{54.5}$ ergs. While the distribution of  GRB parameters may emerge from a variety of considerations, we may be certain that at each stage of every burst, the distribution of viewing angles is isotropically distributed, and that the observers see a wide distribution of isotropic equivalent energies.\footnote{It can safely be assumed that GRB are randomly oriented and that we at Earth represent a fair distribution of observers  observing a single GRB.}

Now consider the X-ray tails (XRTs) of short GRB. Typically the spectral peak energy $h\nu_{peak}$ is of order 10 KeV, about $10^{-2.5}$ of the highest values of $h\nu_{peak}$ among GRB.  So, in the proposed model, the typical value of the Doppler factor $D_{XRT}$ at the stage at which the XRT  is observed is {\bf $(1-\beta)/(1-\beta cos\theta_{min} )\simeq (\Gamma\theta_{min})^{-2}\sim 10^{-2.5}$} and the typical angular separation between line of sight and emitting material is for soft X-ray tails   therefore  $\theta_{min} \sim 10^{1.25}/\Gamma_{XRT}$, and the opening angle of the jet, $\theta_o$,  is  comparable or even higher than $\theta_{min}$. 

Now consider the opening  angle of the early hard stage $\gamma$-radiation. One reason the SGRB may be appear under-energetic  relative to the Amati relation, is that the opening angle of the early, hard $\gamma$-radiation $1/\Gamma$ exceeds  $\theta_o$, so that the flux is diluted by the factor $[1-cos\theta_o]/[1-cos\Gamma^{-1}]$. It would then follow that $1/\Gamma   > \theta_o \gtrsim 10^{1.25}/\Gamma_{XRT}$. 

A second possible reason is that the viewing angle explanation for the Amati relation assumes that all GRB have standard energy outputs. At the early  hard stage of the GRB,  however, only a small fraction  $\Gamma/\Gamma_f$, where $\Gamma_f$ is assumed to be the final Lorentz factor, has been reflected off the baryonic shell. This may be the reason they seem under-energetic relative to the Amati relation. As short GRBs are about two orders of magnitude less energetic at a given $h\nu_{peak}$ than long GRBs  (Shahmoradi and Nemiroff, 2011), this reason (taken alone) would imply that the average value of $\Gamma$  is about two orders of magnitude below the value for long GRBs. Here I have used the fact that radiation that accelerates baryons to ultrarelativistic energies puts about half its energy into the baryons and keeps the other half, assuming it scatters isotropically in the frame of the accelerating baryons,  Both of these arguments  suggest  that the Lorentz factor of the short GRB, at the stage when the hard $\gamma$-rays are radiated, is one to two orders of magnitude less than the final value $\Gamma_F \sim 300$, i.e. between 3 and 30. So the hard $\gamma$-ray beam from short GRBs may in fact be the opening angle convoluted by a $1/\Gamma$ cone as large as 0.03 to 0.3 radian, and a significant fraction of NS merger events may be accompanied by a detectable short GRB.  The value of $\Gamma_F$ is not known accurately, and this makes a more confident quantitative prediction difficult.

   A combined gravitational wave signal and short GRB detections would provide a simple test of the above kinematic model for the difference between short and long GRB, assuming that short GRB are  made by coalescing neutron stars and that the axis of the GRB jet is aligned with the  orbital angular momentum axis of the neutron star binary.   
  Identifying the host galaxy of the short GRB, and accurately determining the masses of the coalescing neutron stars\footnote{The gravitational wave signal is turned off by tidal disruption of the neutron star(s). Unless the equation of state is known exactly, the exact orbital radius at which a neutron star is tidally disrupted by a second NS  (or by a black hole) is also uncertain, and hence  there may be some uncertainty in determining the masses of the neutron stars exactly given only the  orbital frequency and orbital contraction rate.  However it can probably be assumed that the NS mass 
is 1.4 $M_{\odot}$ to 10 percent accuracy.} and the strength of the gravitational wave signal, which is viewing angle dependent,\footnote{A binary neutron star viewed down the orbital rotation axis sees the mass quadrupole moment changing in two directions, whereas there is only one such direction if the orbital plane is seen edge on.}   would in essence measure the viewing angle.  Alternatively, with three separate gravitational wave antennae, each with a different orientation, the polarization of the wave should in principle be measurable and should  contain enough information to determine the viewing angle.  All else failing, the fraction of   GW events (eventually seen by advanced GW detectors) that are accompanied by short GRB would provide information about the opening angles of the latter. With a statistically viable  sample,  the hypothesis that short GRB have larger opening angles predict a broader angular distribution than the opening angles of long GRBs.  
  
  More generally, a large sample of NS merger events would display a larger variety of  GRB.  For example, a NS merger event that is observed by the occasional observer with an unusually  small offset angle $\theta$ is predicted by this model to see brighter, harder extended emission  than typical short GRB X-ray tails, because its emission would begin to decline and soften for kinematic reasons only at a later stage of the GRB, when $\Gamma$ finally exceeds $1/\theta$. It may be that the GRB 060614  (Gehrels et al, 2006) is an  example of such a burst.   If there is negligible viewer  offset, and the short GRB is nevertheless visible because of little or no  intervening material, then the spectrum should be extremely hard near its peak. It may be that GRB 090510, which had an extremely low afterglow and a record high $h\nu_{peak}$ of $\sim 4$ MeV,  is an example of such a short GRB (Eichler, 2014). Note that if GRB 090510 had been viewed at an angle $\theta$ where $h\nu_{peak}$ was the more typical value for long GRB of 200 KeV, (i.e. $(1-\beta)/(1-\beta cos\theta) \sim 200KeV/4MeV=1/20$ then its apparent duration might have appeared 20 times longer, and, although showing some signatures of short GRB, would have appeared long in duration as did GRB 060614.
  Correlating their properties   with measured  viewing angle could help unravel the variety of  GRBs.
  
\section{GRB 170817A} It was { announced shortly} after this paper was originally submitted, that  GRB170817A had a spectral peak of only $\sim 200$ keV, making it unusually soft relative to most hard, short GRB, and that its isotropic equivalent energy was by the most liberal estimate (assumption of Comptonized spectrum and 1 $\sigma$ above  the most probable value) $ 10^{47} $ ergs (Goldstein et al, 2017). This is fully consistent with its being observed from a relatively large angle from the motion of the surface of last interaction, as one would expect from a nearby source  simply because most observers are at large angle.\footnote{It should be clear that such observers would fail to detect distant sources} About five orders of magnitude dimmer (in $W_{iso}$) than long GRB with the same spectral peak (Amati 2002), and about three orders of magnitude dimmer than short GRB with the same spectral peak (Ghirlanda et al, 2009), GRB 170817A is clearly an outlier to the correlations noted by these authors. We may therefore argue that our line of sight is offset from the jet enough that the energy $W_{iso}$ scales more like $(h\nu_{peak})^3$, as would be expected for a pencil beam viewed off axis, as opposed to the Amati relation $W_{iso} \propto (h\nu_{peak})^2$, as would be expected for an observer within the angular perimeter of the jet or just outside of it.  As most of the emission came within 256 ms (Goldstein et al, 2017) the peak luminosity may have been $\gtrsim 4 \times 10^{47}$ erg s$^{-1}$. This is at most  $10^{-6}$ times the $W_{iso}$ of the short GRB with the highest values of  $h\nu_{peak}$, while the peak spectrum $180 keV$, with $\sim 30\%$ uncertainty, is $\simeq 22$ times less than the hardest short GRBs,  e.g. GRB090510. As luminosity scales as $(h\nu_{peak})^4$, this is only slightly below  the expected value of $22^{-4}\simeq 4 \times 10^{-6}$ below the brightest short GRB. Given all the uncertainties, this  is not in  bad agreement. 

The spectrum softened from a peak of $\sim 180$ keV to about 30 keV during 2 seconds. This is consistent with acceleration to a Lorentz factor in the later stages that is $\sim 6$ times the value at the beginning. 

  There is ejected matter observed, as anticipated by many authors (Lattimer and Schramm 1974; Eichler, et al 1989; Paczynski, 1990,  Mezsaros and Rees, 1992;  Woosley and Baron, 1992; Levinson and Eichler, 1993)   following GRB 170817A.  { All these theoretical papers predict an  ejected mass of} at least $10^{-3} M_{\odot}$, and this appears to be consistent with observations of the kilonova associated with the event (Abbott et al, 2017). {The mass ejection} should commence before or within a couple of orbits (milliseconds or less) after the end of the GW signal, which occurs once at least one of the NS is tidally disrupted and converted to a hot accretion disk. The wind, apart from  tidal ejection, should be driven by neutrino heating of the baryonic material, should  therefore be omnidirectional, and should  commence before the GRB is detected 1.7  seconds later (in observer time of course).   Given the timescale of the burst, the ejected mass should have traveled no further than $\sim 2$ light seconds,  At this distance the ejecta would still have a huge optical depth  - $\tau_{opt}\gtrsim 10^{-3}M_{\odot}/(2\, \rm ls)^2 \gtrsim 10^6$  The hypothesis that the surface of last interaction is moving at an offset angle to the line of sight is thus well motivated. It could be that the jet broke through the front of the wind within those two seconds, but then the swept up wind matter at the front of the jet would then have had to clear away in essentially  its entirety in order to be optically thin. 
  
    Alternatively, the GRB may have been powered by neutrino-pair annihilation that happens ahead of the wind. In this case the 1.7 s delay between the end of the GW signal and the GRB would be due to a kinematic delay of $(1-\beta  cos\theta)R/c$,  where R is the distance from the central object at which  the emission is released.  
  
  The above interpretations of the observations are not meant to rule out alternatives, but rather to show how simultaneous observations of GRB and GW events, preferably with a good estimate of the viewing angle, could make available new information about GRB.  For example, if the distribution of delays $t_d$ in a   statistically viable sample were found to be  broad yet to have a sharp, finite lower limit, $t_{min}$, it might be evidence that an event horizon is a necessary ingredient for a GRB, especially if the additional delay $t _d - t_{min}$ were correlated with $cos\theta$. The natural interpretation of $t_{min}$ would be that it is the time it takes the neutron disk to shed enough angular momentum to form a black hole.  Moreover if NS-BH mergers, in contrast to NS-NS mergers,  show no finite $t_{min}$,  it would also be evidence that the delay in NS-NS mergers is partly the interval required for the event horizon to form.

I acknowledge support from  the Israel - U.S. Binational  Science Foundation, the Israel Science Foundation, and the Joan and Robert Arnow Chair of Theoretical Astrophysics. I thank Prof. R. Brustein for a helpful discussion, and  Y. Lyubarsky and L. Ziegler for scrutinizing the manuscript.
\bigskip

 \bigskip
 
 \bigskip
 
\centerline{\bf References}

\noindent Abbott B.P. et al, 2017,    astro-ph https://arxiv.org/pdf/1710.05836.pdf

 \noindent Amati, L., Mon. Not. R. Astron. Soc. 372, 233 (2006). doi:10.1111/j.1365-2966.2006.10840

\noindent Amati, L., Frontera, F., Tavani, M., in’t Zand, J.J.M., Antonelli, A., Costa, E., Feroci, M., Guidorzi, C., Heise, J., Masetti, N., Montanari, E., Nicastro, L., Palazzi, E., Pian, E., Piro, L., Soffitta, P.: Astron. Astrophys. 390, 81 (2002). doi:10.1051/0004-6361:20020722

\noindent Amati, L., Guidorzi, C., Frontera, F., Della Valle, M., Finelli, F., Landi, R., Montanari, E.: Mon. Not. R. Astron. Soc. 391, 577 (2008). arXiv:0805.0377. doi:10.1111/j.1365-2966.2008.13943.x

\noindent Amati, L., Frontera, F., Guidorzi, C.: Astron. Astrophys. 508, 173 (2009). arXiv:0907.0384. doi:10.1051/0004-6361/200912788

\noindent  Abbott B.P. et al. (LIGO Scientific Collaboration and Virgo Collaboration) 2017, Phys. Rev. Lett. 119, 161101

\noindent  Duncan, R.C., Shapiro, S. and Wasserman, I., 1986, ApJ 309, 141

\noindent  Eichler, D. and Manis, H. 2007, ApJ,, 669,  L65 

\noindent  Eichler, D. and Manis, H., 2008, ApJ, 689, L85,

\noindent  Eichler, D., Guetta, D. and Manis, H.  2009,  ApJ., 690, L61

\noindent  Eichler, D. 2014, Ap J. 787, L32

\noindent  Gehrels, N., et al. 2006, Nature, 444, 1044

\noindent Goldstein, A. et al, 2017, ApJ 848, L14

\noindent  Levinson, A. and Eichler, D. 1993, 418, 386

\noindent   Li, L.  and Paczyński B., 1998, ApJ 507,  L59  doi:10.1086/311680. arXiv:astro-ph/9807272

\noindent  Mazets, E. P., et al. 2002, Konus Catalog of Short GRBs (St. Petersburg: Ioffe \\ 
LEA), http://www.ioffe.ru/LEA/shortGRBs/Catalog/

\noindent  Meszaros, P. and Rees, M.J. 1992, ApJ., 397, 570

\noindent  Norris, J. P., et al. 1996, ApJ, 459, 393

\noindent  Norris, J.P.,  Marani, G.F.,  Bonnell, J.T.   2000,  ApJ,  534,  248   

\noindent  Norris, J.P.,   Bonnell, J.T.   2006,  ApJ, 643, 266 

\noindent Paczynski, B. 1990,   1990, ApJ, 363, 218
 
\noindent (Shahmoradi, A.  Nemiroff,  R. 2011, MNRAS, 411, 1843 

\noindent Woosley, S.E. and Baron, E., 1992, ApJ., 391, 228
\end{document}